\documentclass[prd,showpacs,showkeys,floatfix,twocolumn,amsmath,amssymb,floatfix]{revtex4}
\usepackage{graphicx,color,dcolumn,booktabs,bm}
\usepackage{longtable,lscape}
\usepackage{amssymb}
\usepackage{indentfirst}
\usepackage{epsfig}
\usepackage{feynmf}   %{feynmp}
\usepackage{epstopdf}   %{feynmp}
\usepackage{slashed}  %for Feynman symbols
\usepackage{cases}
\usepackage{color}
\usepackage{multirow}
\usepackage{graphicx,color,dcolumn,booktabs,bm}
\usepackage[colorlinks, citecolor=blue,anchorcolor=red,menucolor=red, linkcolor=red,filecolor=red,runcolor=red,urlcolor=blue,frenchlinks=red]{hyperref}

\begin{document}

\title{A QCD sum rule calculation for $P$-wave bottom baryons}

\author{Qiang Mao$^{1,2}$}
\author{Hua-Xing Chen$^{2,3}$}
\email{hxchen@buaa.edu.cn}
\author{Wei Chen$^4$}
\email{wec053@mail.usask.ca}
\author{Atsushi Hosaka$^{5,6}$}
\email{hosaka@rcnp.osaka-u.ac.jp}
\author{Xiang Liu$^{7,8}$}
\email{xiangliu@lzu.edu.cn}
\author{Shi-Lin Zhu$^{9,10,11}$}
\email{zhusl@pku.edu.cn}
\affiliation{
$^1$Department of Electrical and Electronic Engineering, Suzhou University, Suzhou 234000, China\\
$^2$School of Physics and Nuclear Energy Engineering and International Research Center for Nuclei and Particles in the Cosmos, Beihang University, Beijing 100191, China \\
$^3$INFN Laboratori Nazionali di Frascati, Via E. Fermi 40, I-00044 Frascati, Italy \\
$^4$Department of Physics and Engineering Physics, University of Saskatchewan, Saskatoon, Saskatchewan, S7N 5E2, Canada \\
$^5$Research Center for Nuclear Physics, Osaka University, Ibaraki 567--0047, Japan \\
$^6$J-PARC Branch, KEK Theory Center, Institute of Particle and Nuclear Studies, KEK, Tokai, Ibaraki 319-1106, Japan \\
$^7$School of Physical Science and Technology, Lanzhou University, Lanzhou 730000, China\\
$^8$Research Center for Hadron and CSR Physics, Lanzhou University and Institute of Modern Physics of CAS, Lanzhou 730000, China\\
$^9$School of Physics and State Key Laboratory of Nuclear Physics and Technology, Peking University, Beijing 100871, China \\
$^{10}$Collaborative Innovation Center of Quantum Matter, Beijing 100871, China \\
$^{11}$Center of High Energy Physics, Peking University, Beijing 100871, China}

\begin{abstract}
We study the $P$-wave bottom baryons using the method of QCD sum rule and heavy quark effective theory.
Our results suggest that $\Lambda_b(5912)^0$ and $\Lambda_b(5920)^0$ can be well described by the baryon doublet $[\mathbf{\bar 3}_F, 1, 1, \rho]$, and they belong to the $SU(3)$ $\mathbf{\bar 3}_F$ multiplets of $J^P=1/2^-$ and $3/2^-$. Their $SU(3)$ flavor partners, $\Xi_b(1/2^-)$ and $\Xi_b(3/2^-)$, have masses $6.06 \pm 0.13$ GeV and $6.07 \pm 0.13$ GeV, respectively, with mass splitting $9 \pm 4$ MeV. The results obtained using baryon doublet $[\mathbf{\bar 3}_F, 1, 0, \lambda]$ are similar and also consistent with the experimental data. We also study the $SU(3)$ $\mathbf{6}_F$ multiplets by using the baryon multiplets $[\mathbf{6}_F, 0, 1, \lambda]$, $[\mathbf{6}_F, 1, 0, \rho]$ and $[\mathbf{6}_F, 2, 1, \lambda]$, and our results suggest that the $P$-wave bottom baryons $\Sigma_b$, $\Xi^\prime_b$ and $\Omega_b$ have (averaged) masses about 6.0 GeV, 6.2 GeV and 6.4 GeV, respectively.
\end{abstract}

\maketitle

\section{Introduction}
\label{sec:intro}

Recent new observations of hadrons are in many cases associated with the presence of heavy quarks.
In this regard, a systematic study of heavy baryons is important.
So far there are two $\Lambda_b$ resonances, $\Lambda_b(5912)^0$ and $\Lambda_b(5920)^0$,
whose quantum numbers are $J^P = 1/2^-$ and $3/2^-$, respectively~\cite{pdg}. They are both $P$-wave bottom baryons, which were studied by Capstick and Isgur in 1986 using the relativistic quark model~\cite{Capstick:1986bm}, and their predicted masses are in very good agreement with the LHCb and CDF results~\cite{Aaij:2012da,Aaltonen:2013tta}. In the following 30 years, various phenomenological models have been applied to study these excited bottom baryons, including the constituent quark model~\cite{Garcilazo:2007eh,Ortega:2012cx,Yoshida:2015tia}, the relativistic quark model~\cite{Ebert:2007nw}, the color hyperfine interaction~\cite{Karliner:2008sv,Karliner:2015ema},
the heavy quark effective theory~\cite{Roberts:2007ni}, the chiral unitary model~\cite{GarciaRecio:2012db,Xiao:2013yca,Liang:2014eba,Torres-Rincon:2014ffa,Lu:2014ina,Garcia-Recio:2015jsa},
the heavy meson effective theory with $1/M$ corrections~\cite{Yasui:2013iga}, the heavy quark symmetry in multihadron systems~\cite{Yamaguchi:2014era},
the quark pair creation model~\cite{Mu:2014iaa}, the Faddeev approach~\cite{Valcarce:2014fma}, the relativistic flux tube model~\cite{Chen:2014nyo}, and
the Lattice QCD~\cite{Burch:2015pka}; see a recent review article for more details~\cite{Crede:2013sze}.

In this paper we shall study $P$-wave bottom baryons using the method of QCD sum rule~\cite{Shifman:1978bx,Reinders:1984sr} in the
framework of heavy quark effective theory (HQET)~\cite{Grinstein:1990mj,Eichten:1989zv,Falk:1990yz}, which has
been successful applied to study heavy mesons containing a single heavy
quark~\cite{Bagan:1991sg,Neubert:1991sp,Neubert:1993mb,Broadhurst:1991fc,Ball:1993xv,Huang:1994zj,Dai:1996yw,Dai:1993kt,Dai:1996qx,Colangelo:1998ga,Dai:2003yg,Zhou:2014ytp},
and heavy baryons also containing a single heavy
quark~\cite{Shuryak:1981fza,Grozin:1992td,Bagan:1993ii,Dai:1995bc,Dai:1996xv,Groote:1996em,Zhu:2000py,Lee:2000tb,Huang:2000tn,Wang:2003zp,Liu:2007fg}.
There are also some studies using the method of QCD sum rules but not in HQET~\cite{Bagan:1992tp,Bagan:1991sc,Duraes:2007te,Wang:2007sqa}.
During the calculations, we shall take the ${\mathcal O}(1/m_Q)$ corrections ($m_Q$ is the heavy quark mass) into account, and extract the chromomagnetic splitting.

Until now, $\Lambda_b(5912)^0$ and $\Lambda_b(5920)^0$ are the only two $P$-wave bottom baryons well observed in the experiments~\cite{pdg,Aaij:2012da,Aaltonen:2013tta}. This is quite different from charmed baryons~\cite{pdg,Cheng:2015rra}: the four $P$-wave charmed baryons, $\Lambda_c(2595)$, $\Lambda_c(2625)$, $\Xi_c(2790)$ and $\Xi_c(2815)$, are well observed and complete two $SU(3)$ $\mathbf{\bar3}$
multiplets; there are some other candidates, like $\Sigma_c(2800)$, $\Xi_c(2980)$ and $\Xi_c(3080)$, etc.
However, future experiments, such as LHCb and BelleII, are able to search for more bottom baryons, as well as more charmed baryons. Hence, it is a suitable time to carry out this study and write this paper, after we have studied $P$-wave charmed baryons in Ref.~\cite{Chen:2015kpa} which already contains much information.
In this paper, we shall use the same procedures and study $P$-wave bottom baryons.

This paper is organized as follows. In Sec.~\ref{sec:sumrule}, we briefly introduce
the interpolating fields for $P$-wave bottom baryons, and briefly introduce how to use them to
perform QCD sum rule analyses at the leading order and at the ${\mathcal O}(1/m_Q)$
order. The procedures are identical to what we have done in Ref.~\cite{Chen:2015kpa}
for $P$-wave charmed baryons.
In this paper we just need to redo numerical analyses using the $bottom$ quark mass instead of the $charm$ quark mass.
These analyses are done in Sec.~\ref{sec:numerical}, where we use the bottom baryon doublet $[\Lambda_b, 1, 1, \rho]$ as an example.
In Sec.~\ref{sec:summary} we discuss the results and offer a summary.

\section{QCD Sum Rule Analyses}
\label{sec:sumrule}

\begin{figure*}[hbt]
\begin{center}
\scalebox{0.6}{\includegraphics{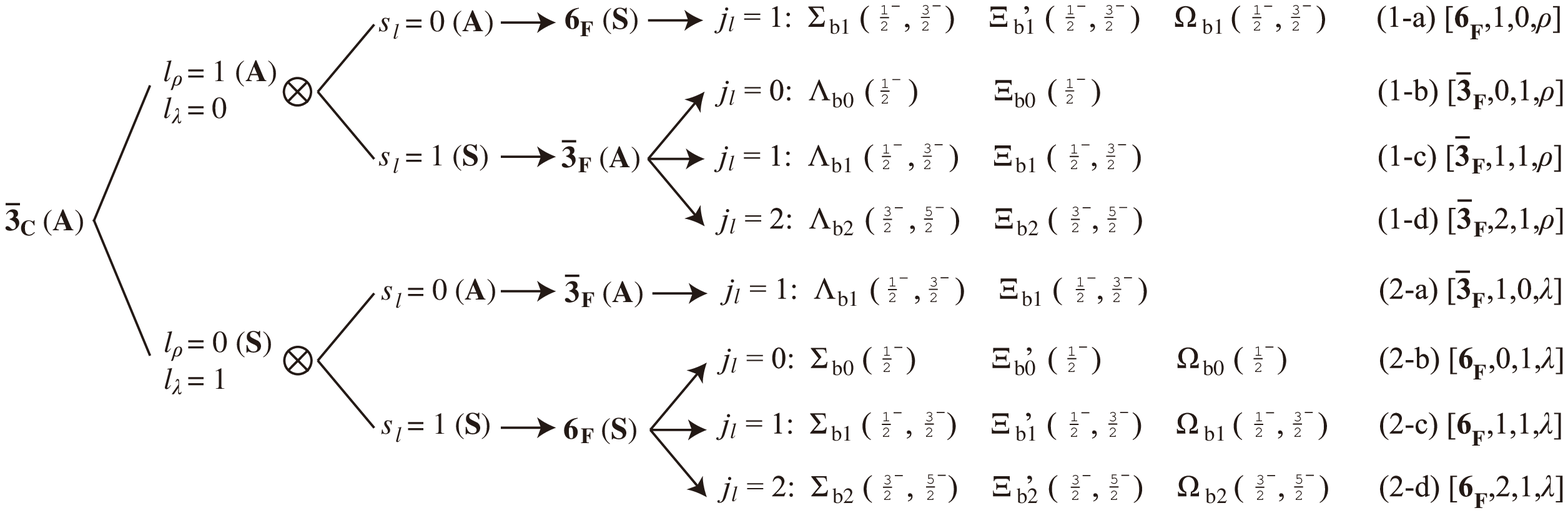}}
\end{center}
\caption{The notations for $P$-wave bottom baryons. See Fig.~2 of Ref.~\cite{Chen:2015kpa} and discussions therein for details.
\label{fig:pwave}}
\end{figure*}

$P$-wave heavy baryons have been systematically classified in Ref.~\cite{Chen:2007xf}, $P$-wave charmed baryon interpolating fields have been systematically constructed in Ref.~\cite{Chen:2015kpa}, and $P$-wave bottom baryon fields can be easily constructed by replacing the $charm$ quark field by the $bottom$ quark field. The results are briefly shown in Fig.~\ref{fig:pwave}, which is
taken from Ref.~\cite{Chen:2015kpa} and slightly modified. For the notation the reader is advised to consult the discussion in Ref.~\cite{Chen:2015kpa}.

In this paper we also use the same notation, $J^{\alpha_1\cdots\alpha_{j-1/2}}_{j,P,F,j_l,s_l,\rho/\lambda}$, to denote the $P$-wave bottom baryon field belonging to the baryon multiplet $[F, j_l, s_l, \rho/\lambda]$, where $j$, $P$, and $F$ are the total angular momentum, parity and $SU(3)$ flavor representation ($\mathbf{\bar 3}_F$ or $\mathbf{6}_F$) of the bottom baryons; $j_l$ and $s_l$ are the total angular momentum and spin angular momentum of the light components; $\rho$ and $\lambda$ denote $\rho$--type ($l_\rho = 1$ and $l_\lambda = 0$) and $\lambda$--type ($l_\rho = 0$ and $l_\lambda = 1$), respectively, where $l_\rho$ is the orbital angular momentum between the two light quarks, and $l_\lambda$ is the orbital angular momentum between the bottom quark and the two-light-quark system. Their relations are $j_l = l_\lambda \otimes l_\rho \otimes s_l$ and $j = j_l \otimes s_Q = | j_l \pm 1/2 |$, where $s_Q = 1/2$ is the spin of the bottom quark.

The explicit forms of $J^{\alpha_1\cdots\alpha_{j-1/2}}_{j,P,F,j_l,s_l,\rho/\lambda}$ have been given in Eqs.~(8)--(21) of Ref.~\cite{Chen:2015kpa}, but $J^{\alpha_1\alpha_2}_{j,-,\mathbf{\bar3}_F,2,1,\rho}$ and $J^{\alpha_1\alpha_2}_{j,-,\mathbf{6}_F,2,1,\lambda}$ therein contain both $j=3/2$ and $j=5/2$ components. In this paper we give the pure $j=5/2$ fields
\begin{eqnarray}
&& J^{\alpha_1\alpha_2}_{5/2,-,\mathbf{\bar3}_F,2,1,\rho}
\\ \nonumber && = i \epsilon_{abc} \Big ( [\mathcal{D}_t^{\mu} q^{aT}] \mathbb{C} \gamma_t^{\nu} q^b - q^{aT} \mathbb{C} \gamma_t^{\nu} [\mathcal{D}_t^{\mu} q^b] \Big )
\times \Gamma^{\alpha_1\alpha_2,\mu\nu} h_v^c \, ,
\\ && J^{\alpha_1\alpha_2}_{5/2,-,\mathbf{6}_F,2,1,\lambda}
\\ \nonumber && = i \epsilon_{abc} \Big ( [\mathcal{D}_t^{\mu} q^{aT}] \mathbb{C} \gamma_t^{\nu} q^b + q^{aT} \mathbb{C} \gamma_t^{\nu} [\mathcal{D}_t^{\mu} q^b] \Big )
\times \Gamma^{\alpha_1\alpha_2,\mu\nu} h_v^c \, ,
\end{eqnarray}
where $\Gamma^{\alpha_1\alpha_2,\mu\nu}$ is the projection operator:
\begin{eqnarray}
\Gamma^{\alpha\beta,\mu\nu} &=& g_t^{\alpha\mu} g_t^{\beta\nu} + g_t^{\alpha\nu} g_t^{\beta\mu} - {2 \over 15} g_t^{\alpha\beta} g_t^{\mu\nu}
\\ \nonumber && - {1 \over 3} g_t^{\alpha\mu} \gamma_t^{\beta}\gamma_t^{\nu} - {1 \over 3} g_t^{\alpha\nu} \gamma_t^{\beta}\gamma_t^{\mu}
\\ \nonumber && - {1 \over 3} g_t^{\beta\mu} \gamma_t^{\alpha}\gamma_t^{\nu} - {1 \over 3} g_t^{\beta\nu} \gamma_t^{\alpha}\gamma_t^{\mu}
\\ \nonumber && + {1 \over 15} \gamma_t^{\alpha} \gamma_t^{\mu} \gamma_t^{\beta} \gamma_t^{\nu} + {1 \over 15} \gamma_t^{\alpha} \gamma_t^{\nu} \gamma_t^{\beta} \gamma_t^{\mu}
\\ \nonumber && + {1 \over 15} \gamma_t^{\beta} \gamma_t^{\mu} \gamma_t^{\alpha} \gamma_t^{\nu} + {1 \over 15} \gamma_t^{\beta} \gamma_t^{\nu} \gamma_t^{\alpha} \gamma_t^{\mu} \, .
\end{eqnarray}
The notations used in these equations are: $a$, $b$ and $c$ are color indices, and $\epsilon_{abc}$ is the totally antisymmetric tensor; the superscript $T$ represents the transpose of the Dirac indices only; $\mathbb{C}$ is the charge-conjugation operator; $q(x)$ is the light quark field at location $x$, and it can be either $u(x)$ or $d(x)$ or $s(x)$; $h_v(x)$ is the heavy quark field, and we have used the Fierz transformation to move it to the rightmost place; $D^\mu = \partial^\mu - i g A^\mu$ is the gauge-covariant derivative; $v$ is the velocity of the heavy quark; $\gamma_t^\mu = \gamma^\mu - v\!\!\!\slash v^\mu$, $D^\mu_t = D^\mu - (D \cdot v) v^\mu$, and $g_t^{\alpha_1\alpha_2}=g^{\alpha_1\alpha_2} - v^{\alpha_1} v^{\alpha_2}$.

We can use these $P$-wave bottom baryon interpolating fields to perform QCD sum rule analyses. The heavy baryon belonging to the multiplet $[F, j_l, s_l, \rho/\lambda]$ has the mass:
\begin{eqnarray}
m_{j,P,F,j_l,s_l,\rho/\lambda} = m_Q + \overline{\Lambda}_{F,j_l,s_l,\rho/\lambda} + \delta m_{j,P,F,j_l,s_l,\rho/\lambda} \, ,
\label{eq:mass}
\end{eqnarray}
where $m_Q$ is the heavy quark mass, $\overline{\Lambda}_{F,j_l,s_l,\rho/\lambda} = \overline{\Lambda}_{|j_l-1/2|,P,F,j_l,s_l,\rho/\lambda} = \overline{\Lambda}_{j_l+1/2,P,F,j_l,s_l,\rho/\lambda}$ is the sum rule result evaluated at the leading order, and $\delta m_{j,P,F,j_l,s_l,\rho/\lambda}$ is the sum rule result evaluated at the ${\mathcal O}(1/m_Q)$ order:
\begin{eqnarray}
&& \delta m_{j,P,F,j_l,s_l,\rho/\lambda}
\label{eq:masscorrection}
\\ \nonumber && = -\frac{1}{4m_{Q}}(K_{F,j_l,s_l,\rho/\lambda} + d_{M}C_{mag}\Sigma_{F,j_l,s_l,\rho/\lambda} ) \, .
\end{eqnarray}
Here, the coefficient $d_{M} \equiv d_{j,j_{l}}$ is defined to be
\begin{eqnarray}
d_{j_{l}-1/2,j_{l}} &=& 2j_{l}+2\, ,
\\ \nonumber d_{j_{l}+1/2,j_{l}} &=& -2j_{l} \, .
\end{eqnarray}
It is directly related to the baryon mass splitting within the same multiplet, caused by the chromomagnetic interaction. The other coefficient is $C_{mag} (m_{Q}/\mu) = [ \alpha_s(m_Q) / \alpha_s(\mu) ]^{3/\beta_0} \approx 0.8$, where $\beta_0 = 11 - 2 n_f /3$.

The (analytical) formulae for $\overline{\Lambda}_{F,j_l,s_l,\rho/\lambda}(\omega_c, T)$, $K_{F,j_l,s_l,\rho/\lambda}(\omega_c, T)$ and $\Sigma_{F,j_l,s_l,\rho/\lambda}(\omega_c, T)$ have been obtained in Ref.~\cite{Chen:2015kpa}. In this paper we just need to use these equations to redo numerical analyses but using the $bottom$ quark mass. There are two free parameters in these equations: the Borel mass $T$ and the threshold value $\omega_c$. Following the procedures used in Ref.~\cite{Chen:2015kpa}, we can obtain a Borel window $T_{min}<T<T_{max}$ for a fixed threshold value $\omega_c$. We shall not repeat these procedures again. The other parameter, $\omega_c$, can be fixed by requiring that its dependence of the mass prediction be the weakest, which will be done in the next section.

\section{Numerical Analyses}
\label{sec:numerical}

To perform numerical analyses, we use the following values for various condensates and parameters~\cite{Dai:1993kt,Dai:1996yw,Dai:1996qx,Dai:2003yg,pdg,Yang:1993bp,Hwang:1994vp,Narison:2002pw,Gimenez:2005nt,Jamin:2002ev,Ioffe:2002be,Ovchinnikov:1988gk,colangelo}:
%
%%%%%%%%%%%%%%%%%%%%%%%%%%%%%%%%%%%%%%%%%%%%%%%%%%%%%%%%%%%%%%%%%%%%%%%%%%%%%%
\begin{eqnarray}
\nonumber && \langle \bar qq \rangle = \langle \bar uu \rangle = \langle \bar dd \rangle = - (0.24 \pm 0.01)^3 \mbox{ GeV}^3 \, ,
\\ \nonumber && \langle \bar ss \rangle = 0.8 \times \langle\bar qq \rangle \, ,
\\ \nonumber &&\langle g_s^2GG\rangle =(0.48\pm 0.14) \mbox{ GeV}^4\, ,
\\ \label{condensates} && m_s = 0.15 \mbox{ GeV} \, ,
\\
\nonumber && \langle g_s \bar q \sigma G q \rangle = M_0^2 \times \langle \bar qq \rangle\, ,
\\
\nonumber && \langle g_s \bar s \sigma G s \rangle = M_0^2 \times \langle \bar ss \rangle\, ,
\\
\nonumber && M_0^2= 0.8 \mbox{ GeV}^2\, .
\end{eqnarray}
%%%%%%%%%%%%%%%%%%%%%%%%%%%%%%%%%%%%%%%%%%%%%%%%%%%%%%%%%%%%%%%%%%%%%%%%%%%%%%
Besides them, another important parameter is the $bottom$ quark mass, for which we use the $1S$ mass $m_b = 4.66 \pm 0.03$ GeV~\cite{pdg}. We shall see that this value gives reasonable results.

\subsection{Sum Rule Analyses for $\Lambda_b$}

As an example, we use the bottom baryon doublet $[\Lambda_b(\mathbf{\bar 3}_F), 1, 1, \rho]$ to perform QCD sum rule analyses. It contains two bottom baryons, $\Lambda_b(1/2^-)$ and $\Lambda_b(3/2^-)$, and the relevant interpolating field is
\begin{eqnarray}
&& J_{1/2,-,\Lambda_b,1,1,\rho}
\label{eq:current}
\\ \nonumber && = i \epsilon_{abc} \Big ( [\mathcal{D}_t^{\mu} u^{aT}] C \gamma_t^\nu d^b -  u^{aT} C \gamma_t^\nu [\mathcal{D}_t^{\mu} d^b] \Big ) \sigma_t^{\mu\nu} h_v^c \, .
\end{eqnarray}
We can use this current to perform QCD sum rule analyses, and calculate $\overline{\Lambda}_{\Lambda_b,1,1,\rho}(\omega_c, T)$, $K_{\Lambda_b,1,1,\rho}(\omega_c, T)$ and $\Sigma_{\Lambda_b,1,1,\rho}(\omega_c, T)$. This has been done and the results are listed in Eq.~(A6) of Ref.~\cite{Chen:2015kpa}, which can be used to further evaluate masses of $\Lambda_b(1/2^-)$ and $\Lambda_b(3/2^-)$ through Eqs.~(\ref{eq:mass}) and (\ref{eq:masscorrection}), as shown below.

\begin{figure}[h]
\begin{center}
\begin{tabular}{c}
\scalebox{0.6}{\includegraphics{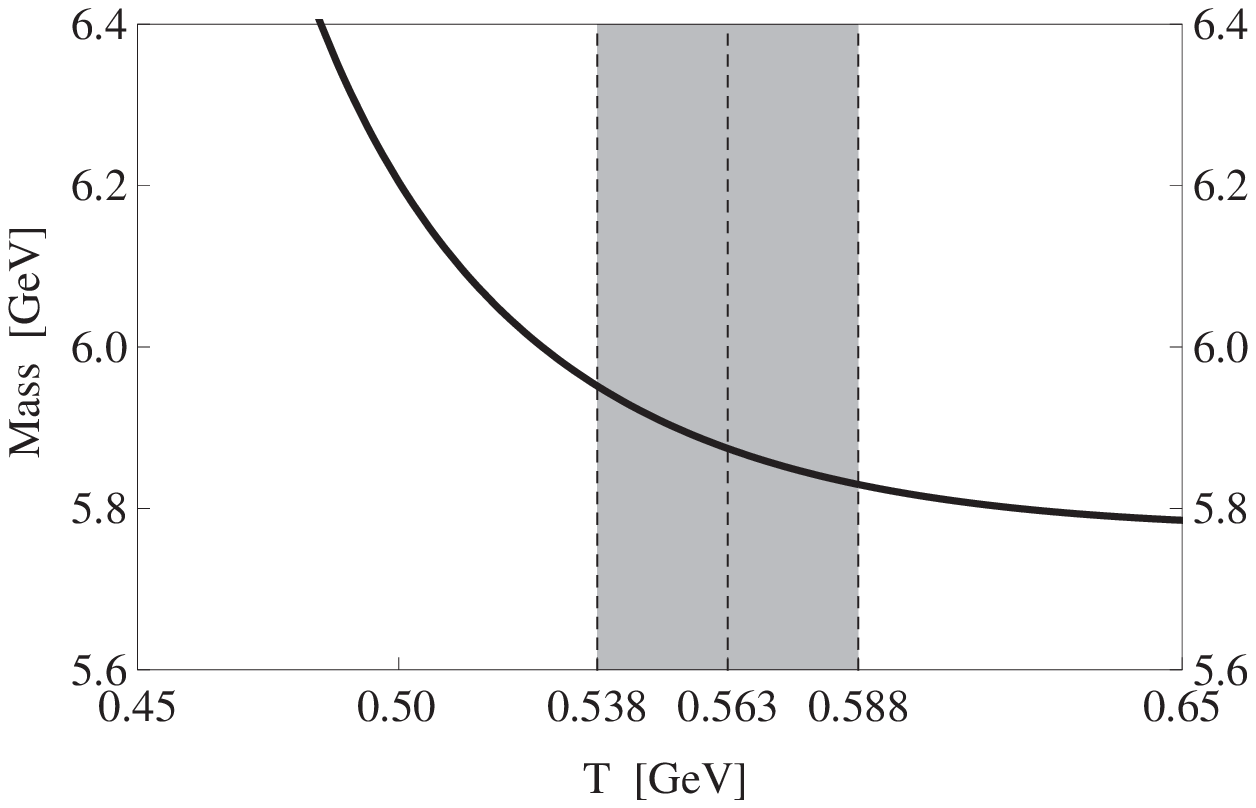}}
\\
\scalebox{0.6}{\includegraphics{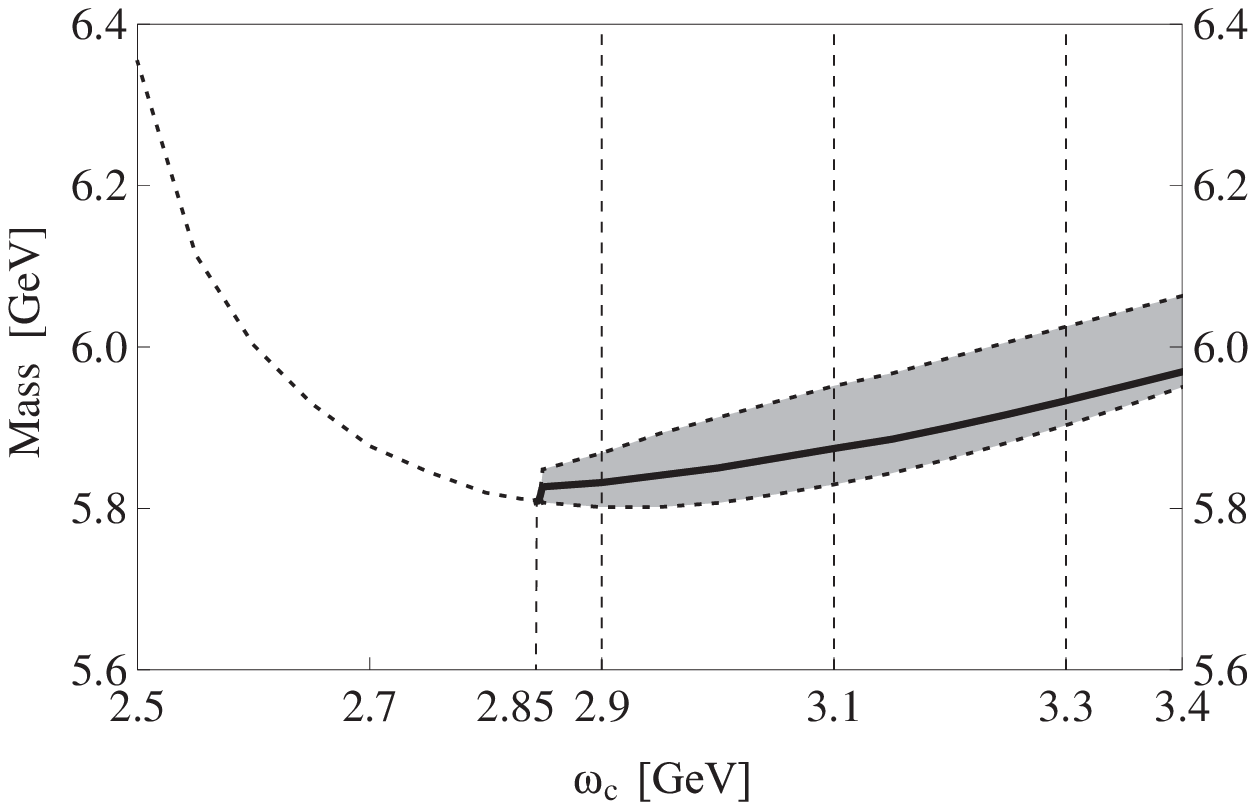}}
\end{tabular}
\caption{Variations of $m_{\Lambda_b(1/2^-)}$ with respect to the Borel mass
$T$ (top) and the threshold value $\omega_c$ (bottom), calculated using the bottom baryon doublet
$[\Lambda_b(\mathbf{\bar 3}_F), 1, 1, \rho]$. In the top panel we take $\omega_c = 3.1$ GeV and the Borel window is
$0.538$ GeV $< T < 0.588$ GeV. In the bottom panel, the upper and lower
bands are obtained by using $T_{\rm min}$ and $T_{\rm max}$, respectively. There are non-vanishing Borel windows as long as $\omega_c \geq 2.85$ GeV, but they quickly vanish around this point. The results for $\omega_c \leq 2.85$ GeV are also shown. For such cases we choose the Borel mass $T$ when the PC (pole contribution, defined in Eq.~(34) of Ref.~\cite{Chen:2015kpa}) is around 20\%, and so this curve connects to the lower band obtained by using $T_{\rm max}$.}
\label{fig:rho3F11}
\end{center}
\end{figure}

\begin{figure}[h]
\begin{center}
\begin{tabular}{c}
\scalebox{0.6}{\includegraphics{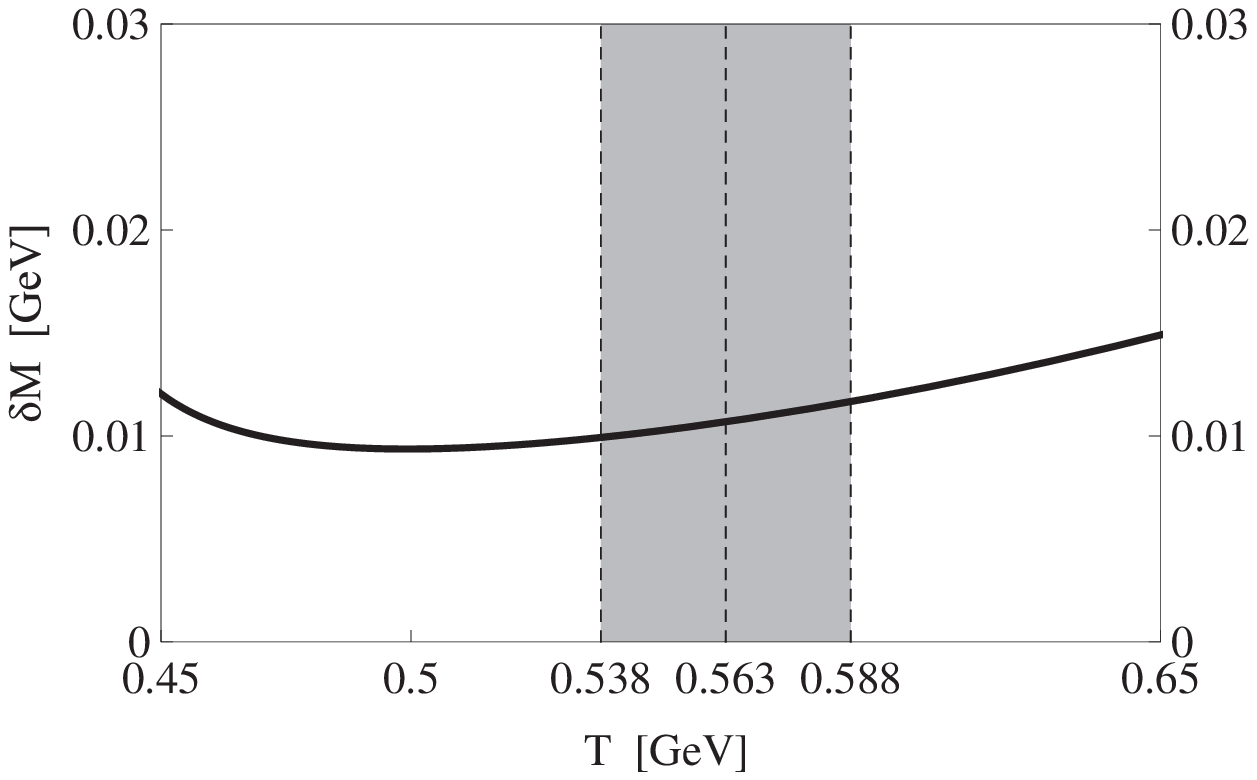}}
\\
\scalebox{0.6}{\includegraphics{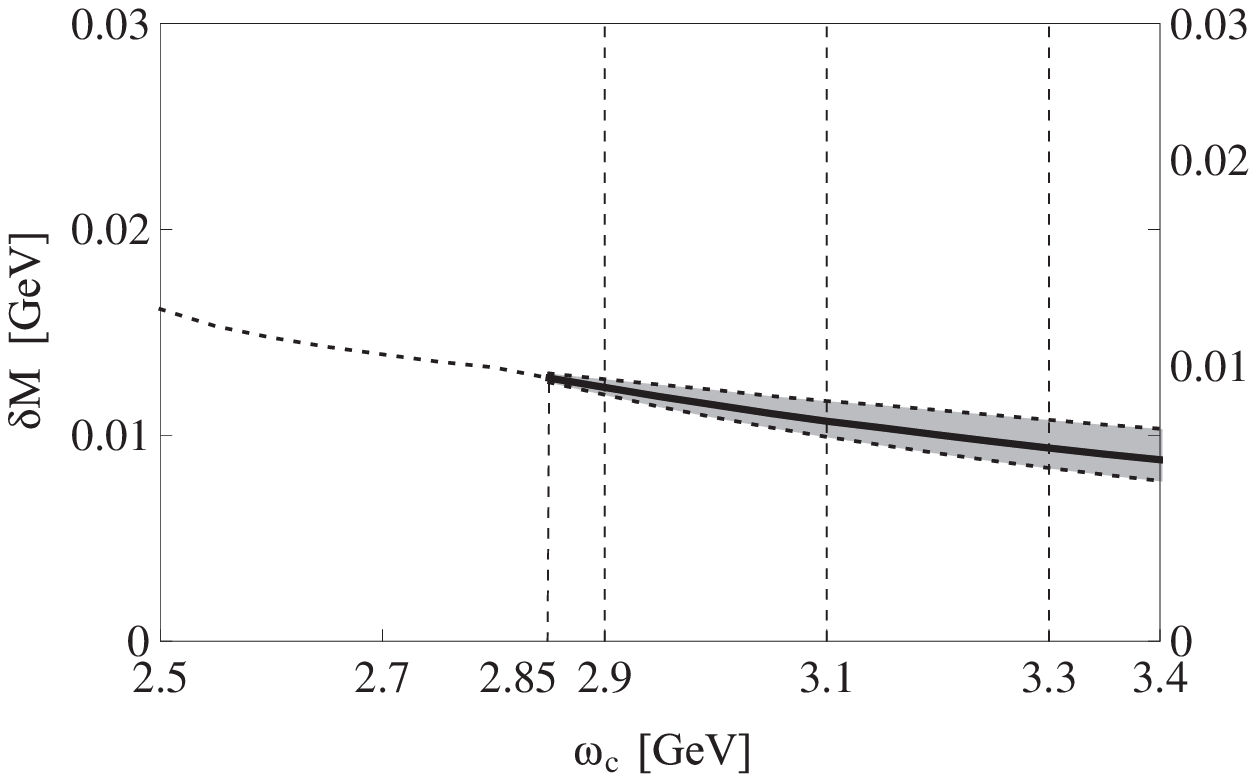}}
\end{tabular}
\caption{Variations of $\Delta m_{[\Lambda_b, 1, 1, \rho]}$ with respect to the Borel mass
$T$ (top) and the threshold value $\omega_c$ (bottom), calculated using the bottom baryon doublet
$[\Lambda_b(\mathbf{\bar 3}_F), 1, 1, \rho]$. In the top panel we take $\omega_c = 3.1$ GeV and the Borel window is
$0.538$ GeV $< T < 0.588$ GeV.}
\label{fig:rho3F11split}
\end{center}
\end{figure}

Firstly, we take $\omega_c = 3.1$ GeV and obtain a Borel window $0.538$ GeV $< T < 0.588$ GeV producing the following numerical results:
\begin{eqnarray}
\nonumber \bar \Lambda_{\Lambda_b,1,1,\rho} &=& 1.17 \mbox{ GeV} \, , \,
\\ K_{\Lambda_b,1,1,\rho} &=& -0.99 \mbox{ GeV}^2 \, , \,
\\ \nonumber \Sigma_{\Lambda_b,1,1,\rho} &=& 0.042 \mbox{ GeV}^{2} \, ,
\end{eqnarray}
whose values correspond to $T=0.563$ GeV and $\omega_c = 3.1$ GeV. Using Eqs.~(\ref{eq:mass}) and (\ref{eq:masscorrection}), we can further obtain
\begin{eqnarray}
\nonumber m_{\Lambda_b(1/2^-)} &=& 5.87 \mbox{ GeV} \, , \,
\\ m_{\Lambda_b(3/2^-)} &=& 5.88 \mbox{ GeV} \, , \,
\\ \nonumber \Delta m_{[\Lambda_b, 1, 1, \rho]} &=& 11 \mbox{ MeV} \, ,
\end{eqnarray}
where $(m_{\Lambda_b(1/2^-)},m_{\Lambda_b(3/2^-)})$ are the masses of the two bottom baryons belonging to the baryon doublet $[\Lambda_b(\mathbf{\bar 3}_F), 1, 1, \rho]$, and $\Delta m_{[\Lambda_b, 1, 1, \rho]} \equiv m_{\Lambda_b(3/2^-)} - m_{\Lambda_b(1/2^-)}$ is their mass splitting. We show variations of $m_{\Lambda_b(1/2^-)}$ and $\Delta m_{[\Lambda_b, 1, 1, \rho]}$ with respect to the Borel mass $T$ in the top panel of Figs.~\ref{fig:rho3F11} and \ref{fig:rho3F11split}.

Secondly, we change the threshold value $\omega_c$ and redo these processes. Our criterion is to require that the $\omega_c$ dependence of
the mass prediction be the weakest. We show variations of $m_{\Lambda_b(1/2^-)}$ and $\Delta m_{[\Lambda_b, 1, 1, \rho]}$
with respect to the threshold value $\omega_c$ in the bottom panel of Figs.~\ref{fig:rho3F11} and \ref{fig:rho3F11split}. We find that there are non-vanishing Borel windows as long as $\omega_c \geq 2.85$ GeV, and
the $\omega_c$ dependence is weak and acceptable in the region $2.9$ GeV$<\omega_c<3.3$ GeV. We note that this dependence is the weakest around $\omega_c \sim 2.9$ GeV, but we prefer to choosing a suitable region where we can know the related uncertainty. The results for $\omega_c \leq 2.85$ GeV are also shown, and for such cases, we choose the Borel mass $T$ when the PC (pole contribution, defined in Eq.~(34) of Ref.~\cite{Chen:2015kpa}) is around 20\%.

Finally, we choose $2.9$ GeV$<\omega_c<3.3$ GeV and $0.538$ GeV $< T < 0.588$ GeV as our working regions, and obtain
the following numerical results for the baryon doublet $[\Lambda_b(\mathbf{\bar 3}_F), 1, 1, \rho]$:
\begin{eqnarray}
\nonumber m_{\Lambda_b(1/2^-)} &=& 5.87 \pm 0.12 \mbox{ GeV} \, , \,
\\ m_{\Lambda_b(3/2^-)} &=& 5.88 \pm 0.11 \mbox{ GeV} \, , \,
\\ \nonumber \Delta m_{[\Lambda_b, 1, 1, \rho]} &=& 11 \pm 4 \mbox{ MeV} \, ,
\end{eqnarray}
whose central values correspond to $T=0.563$ GeV and $\omega_c = 3.1$ GeV, and the
uncertainties are due to the Borel mass $T$, the threshold value $\omega_c$, the bottom quark mass $m_b$ and the
quark and gluon condensates. Our result for the mass splitting within
the same doublet, $\Delta m_{[\Lambda_b, 1, 1, \rho]}$, does not depend much on the charm quark mass and the threshold value, so they are produced quite well,
with much less theoretical uncertainties, compared to their masses, $m_{\Lambda_b(1/2^-)}$ and $m_{\Lambda_b(3/2^-)}$.
These results are consistent with experimental masses of $\Lambda_b(5912)^0$ ($1/2^-$) and $\Lambda_b(5920)^0$ ($J^P=3/2^-$) as well as their difference~\cite{Aaij:2012da,Aaltonen:2013tta}:
\begin{eqnarray}
&& m^{\rm exp}_{\Lambda_b(5912)^0,1/2^-} = 5912.11 \pm 0.13 \pm 0.23 \mbox{ MeV} \, ,
\\ \nonumber && m^{\rm exp}_{\Lambda_b(5920)^0,3/2^-} = 5919.91 \pm 0.09 \pm 0.23 \mbox{ MeV} \, .
\end{eqnarray}
This suggests that the use of the $1S$ mass $m_b = 4.66 \pm 0.03$ GeV~\cite{pdg} is reasonable. While, if we use the $\overline{\rm MS}$ mass $m_b = 4.18 \pm 0.03$ GeV~\cite{pdg}, we would obtain significantly lower masses.

\begin{figure}[hbt]
\begin{center}
\begin{tabular}{c}
\scalebox{0.6}{\includegraphics{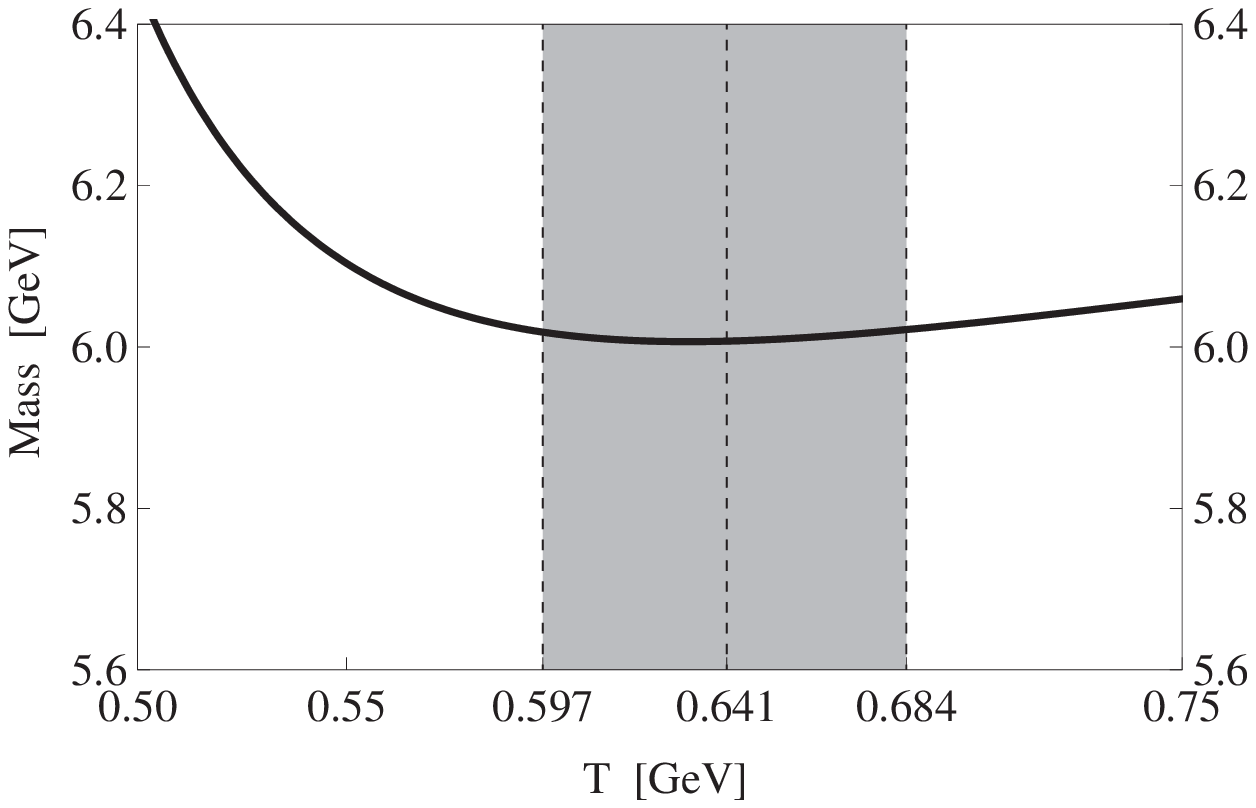}}
\\
\scalebox{0.6}{\includegraphics{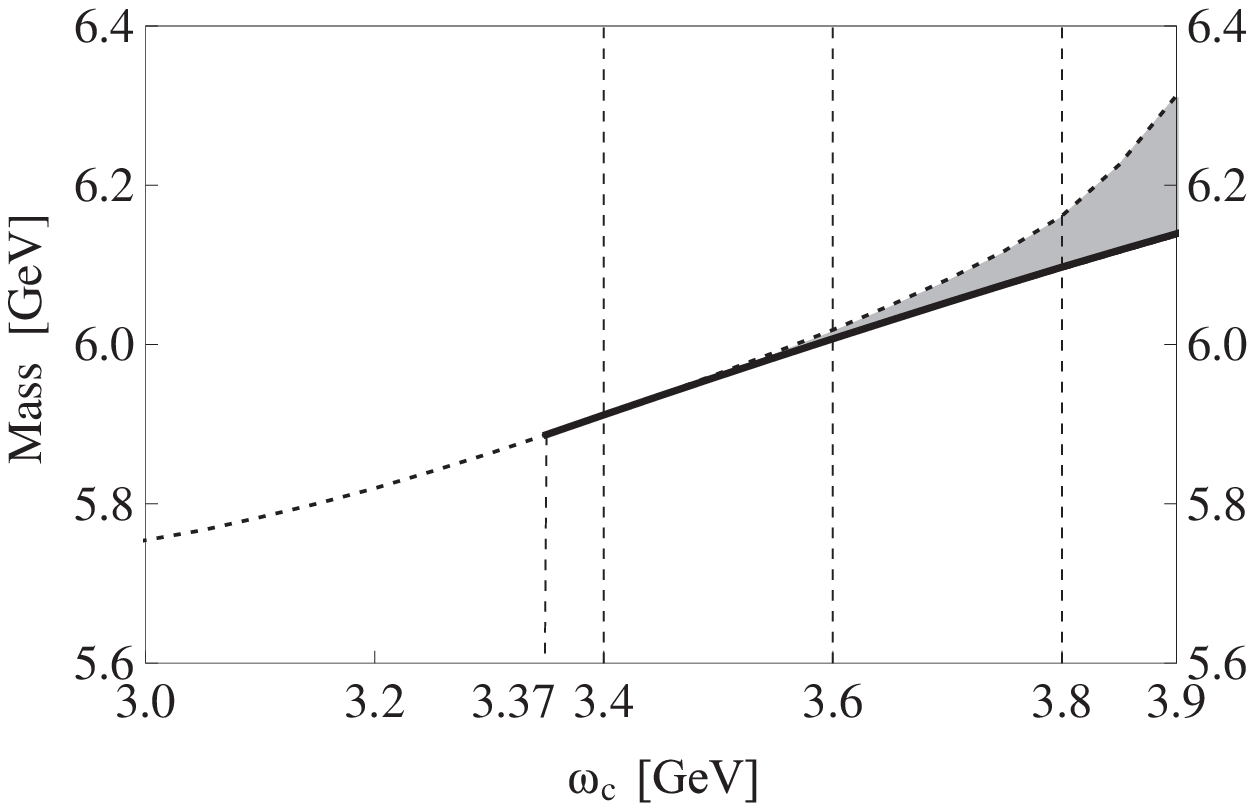}}
\end{tabular}
\caption{Variations of $m_{\Lambda_b(1/2^-)}$ with respect to the Borel mass
$T$ (top) and the threshold value $\omega_c$ (bottom), calculated using the bottom baryon doublet
$[\Lambda_b(\mathbf{\bar 3}_F), 1, 0, \lambda]$. In the top panel we take $\omega_c = 3.6$ GeV and the Borel window is
$0.597$ GeV $< T < 0.684$ GeV and $3.4$ GeV.}
\label{fig:lambda3F10}
\end{center}
\end{figure}

We also use the bottom baryon doublet $[\Lambda_b(\mathbf{\bar 3}_F), 1, 0, \lambda]$ to perform QCD sum rule analyses.
We show the variation of $m_{\Lambda_b(1/2^-)}$
with respect to the Borel mass $T$ and the threshold value $\omega_c$ in Fig.~\ref{fig:lambda3F10}. We find that there are non-vanishing Borel windows as long as $\omega_c \geq 3.37$ GeV, but in this region the $\omega_c$ dependence is not so weak. We still choose $3.4$ GeV $<\omega_c <3.8$ GeV and $0.597$ GeV $< T < 0.684$ GeV as our working regions, but note that the previous results obtained using the doublet $[\Lambda_b(\mathbf{\bar 3}_F), 1, 1, \rho]$ seem to be more stable (reliable), because the $\rho$--mode signal is clearer than the $\lambda$--one. In fact the lower panel of Fig.~\ref{fig:lambda3F10}, which is monotonically increasing, shows that the $\lambda$--mode is not well observed. This does not, however, say that the $\lambda$--mode mass is larger than the $\rho$--mode mass in a reliable manner, though the current choice of the Borel window suggests it. We obtain the following numerical results
\begin{eqnarray}
\nonumber m_{\Lambda_b(1/2^-)} &=& 6.01 \pm 0.11 \mbox{ GeV} \, , \,
\\ m_{\Lambda_b(3/2^-)} &=& 6.01 \pm 0.11 \mbox{ GeV} \, , \,
\\ \nonumber \Delta m_{[\Lambda_b, 1, 0, \lambda]} &=& 5 \pm 2 \mbox{ MeV} \, ,
\end{eqnarray}
whose central values correspond to $T=0.641$ GeV and $\omega_c = 3.6$ GeV. We note again that the mass splitting within the same doublet is reproduced quite well with much less (theoretical) uncertainty than the absolute values. These results are also consistent with masses of $\Lambda_b(5912)^0$ ($1/2^-$) and $\Lambda_b(5920)^0$ ($J^P=3/2^-$) as well as their difference~\cite{Aaij:2012da,Aaltonen:2013tta}.

\subsection{Sum Rule Analysis for $\Sigma_b$, $\Xi_b^{(\prime)}$ and $\Omega_b$}

There are two baryons, $\Lambda_b(5912)^0$ and $\Lambda_b(5920)^0$, which are the only P-wave bottom baryons well observed in experiments~\cite{pdg,Aaij:2012da,Aaltonen:2013tta}, and there are no other well established candidates. To evaluate masses of other baryon multiplets, we follow the previous procedures used for $[\Lambda_b(\mathbf{\bar 3}_F), 1, 1, \rho]$
as well as for charmed baryons~\cite{Chen:2015kpa}:

\begin{enumerate}

\item We use the bottom baryon singlet $[\Sigma_b(\mathbf{6}_F), 0, 1, \lambda]$ to perform QCD sum rule analyses. We show the variation of $m_{\Sigma_b(1/2^-)}$ with respect to the Borel mass $T$ and the threshold value $\omega_c$ in Fig.~\ref{fig:lambda6F01}. We find that there are non-vanishing Borel windows as long as $\omega_c \geq 3.06$ GeV, and the $\omega_c$ dependence is weak and acceptable in the region $3.1$ GeV$<\omega_c<3.5$ GeV. Accordingly, we choose $3.1$ GeV $<\omega_c <3.5$ GeV and $0.527$ GeV $< T < 0.628$ GeV as our working regions, and obtain the following numerical result
    \begin{eqnarray}
    m_{\Sigma_b(1/2^-)} &=& 6.02 \pm 0.12 \mbox{ GeV} \, , \,
    \end{eqnarray}
    where the central value corresponds to $T=0.578$ GeV and $\omega_c = 3.3$ GeV.

\begin{figure}[hbt]
\begin{center}
\begin{tabular}{c}
\scalebox{0.6}{\includegraphics{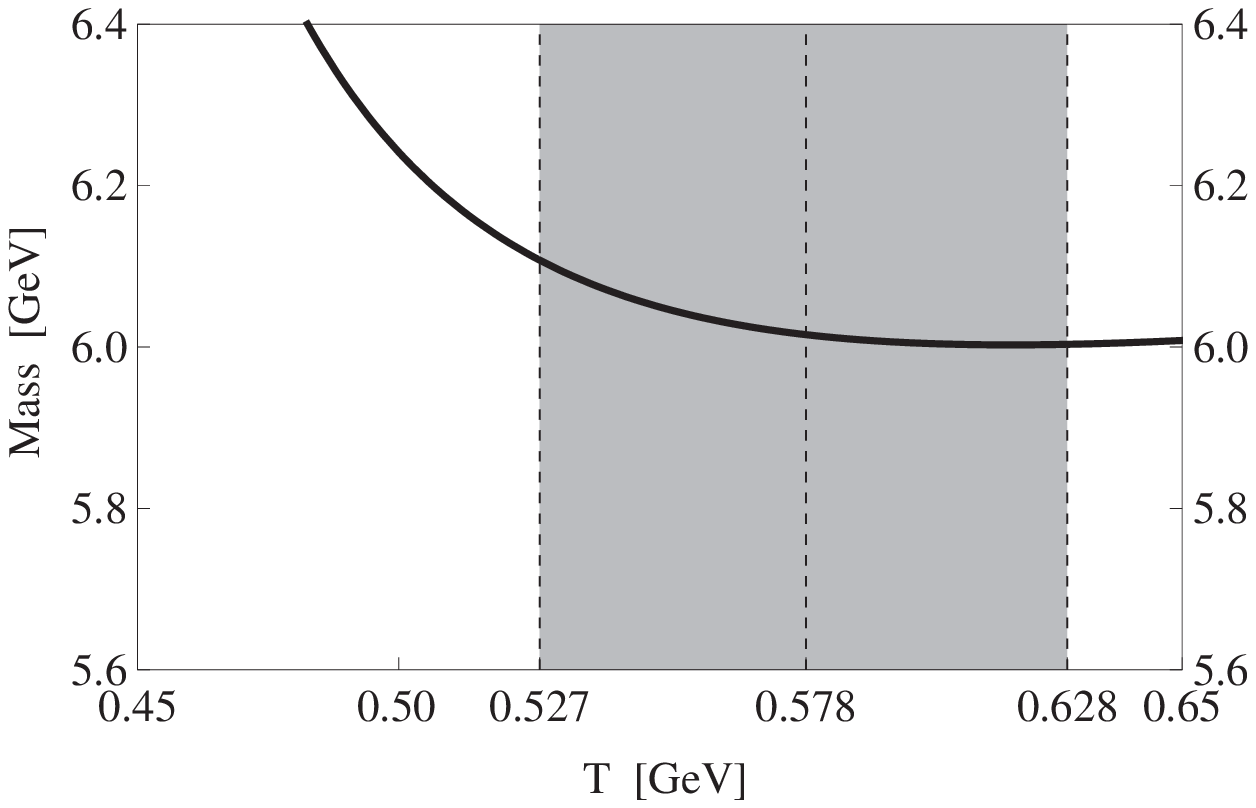}}
\\
\scalebox{0.6}{\includegraphics{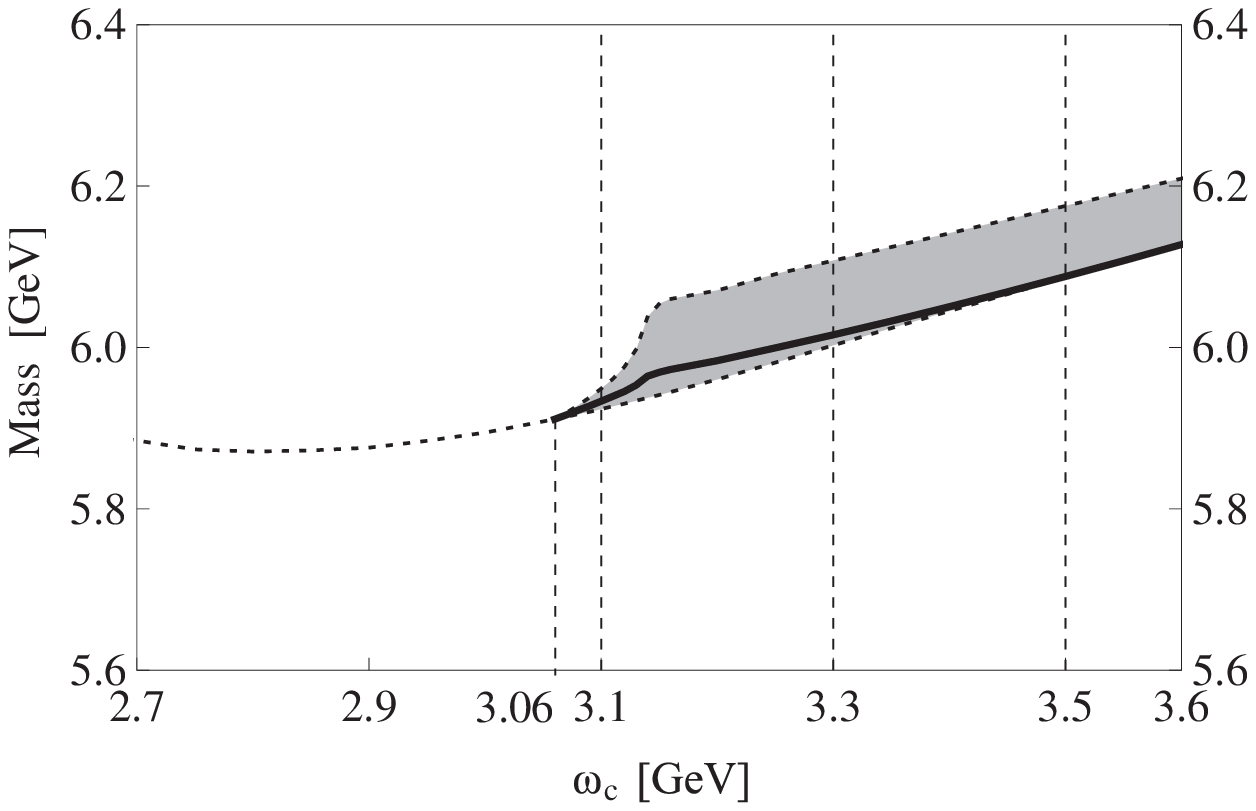}}
\end{tabular}
\caption{Variations of $m_{\Sigma_b(1/2^-)}$ with respect to the Borel mass
$T$ (top) and the threshold value $\omega_c$ (bottom), calculated using the bottom baryon singlet
$[\Sigma_b(\mathbf{6}_F), 0, 1, \lambda]$. In the top panel we take $\omega_c = 3.3$ GeV and the Borel window is
$0.527$ GeV $< T < 0.628$ GeV.}
\label{fig:lambda6F01}
\end{center}
\end{figure}

\item We use the bottom baryon doublet $[\Sigma_b(\mathbf{6}_F), 1, 0, \rho]$ to perform QCD sum rule analyses. We show the variations of $m_{\Sigma_b(1/2^-)}$ with respect to the Borel mass $T$ and the threshold value $\omega_c$ in Fig.~\ref{fig:rho6F10}. We find that there are non-vanishing Borel windows as long as $\omega_c \geq 3.01$ GeV, and the $\omega_c$ dependence is weak and acceptable in the region $3.1$ GeV$<\omega_c<3.5$ GeV. Hence, we choose $3.1$ GeV $<\omega_c <3.5$ GeV and $0.534$ GeV $< T < 0.624$ GeV as our working regions, and obtain the following numerical results
    \begin{eqnarray}
    \nonumber m_{\Sigma_b(1/2^-)} &=& 5.91 \pm 0.14 \mbox{ GeV} \, , \,
    \\ m_{\Sigma_b(3/2^-)} &=& 5.92 \pm 0.14 \mbox{ GeV} \, , \,
    \\ \nonumber \Delta m_{[\Sigma_b, 1, 0, \rho]} &=& 4\pm2 \mbox{ MeV} \, ,
    \end{eqnarray}
    whose central values correspond to $T=0.579$ GeV and $\omega_c = 3.3$ GeV. We note again that the mass splitting within the same doublet is reproduced quite well with much less (theoretical) uncertainty.

\begin{figure}[hbt]
\begin{center}
\begin{tabular}{c}
\scalebox{0.6}{\includegraphics{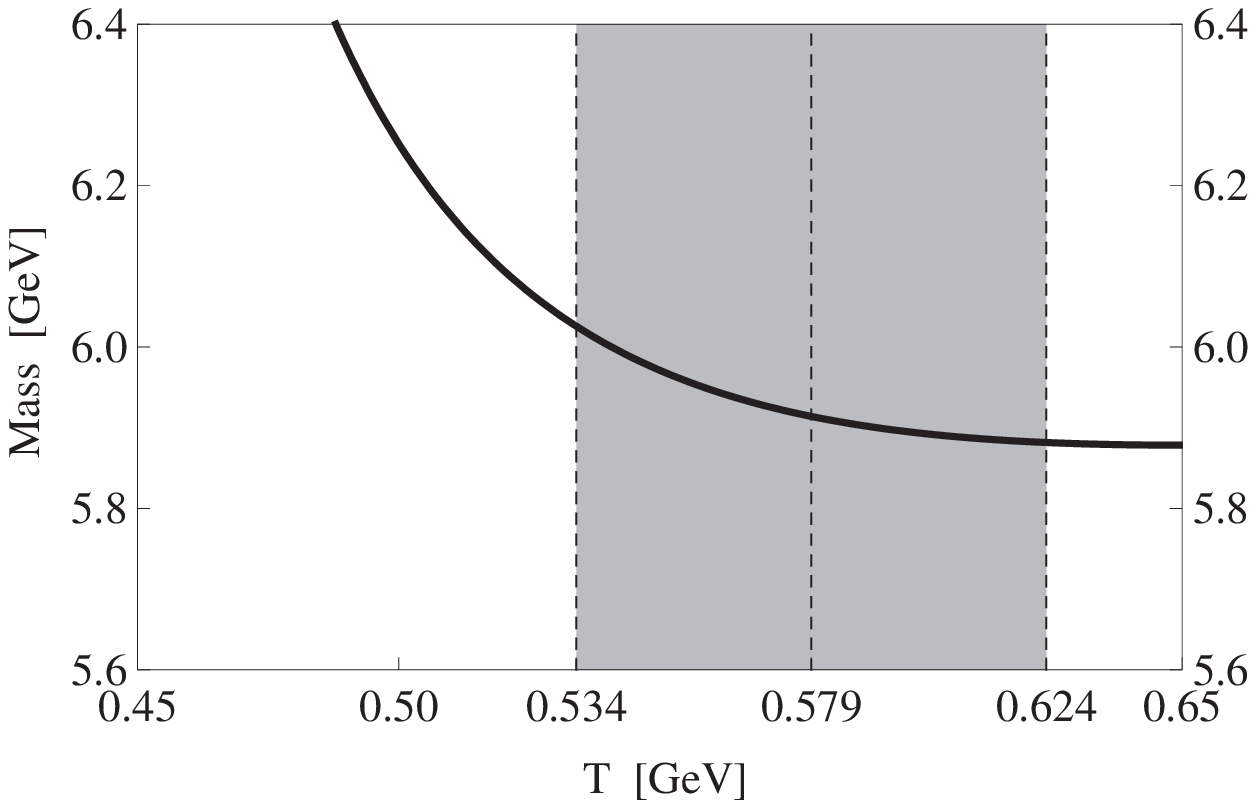}}
\\
\scalebox{0.6}{\includegraphics{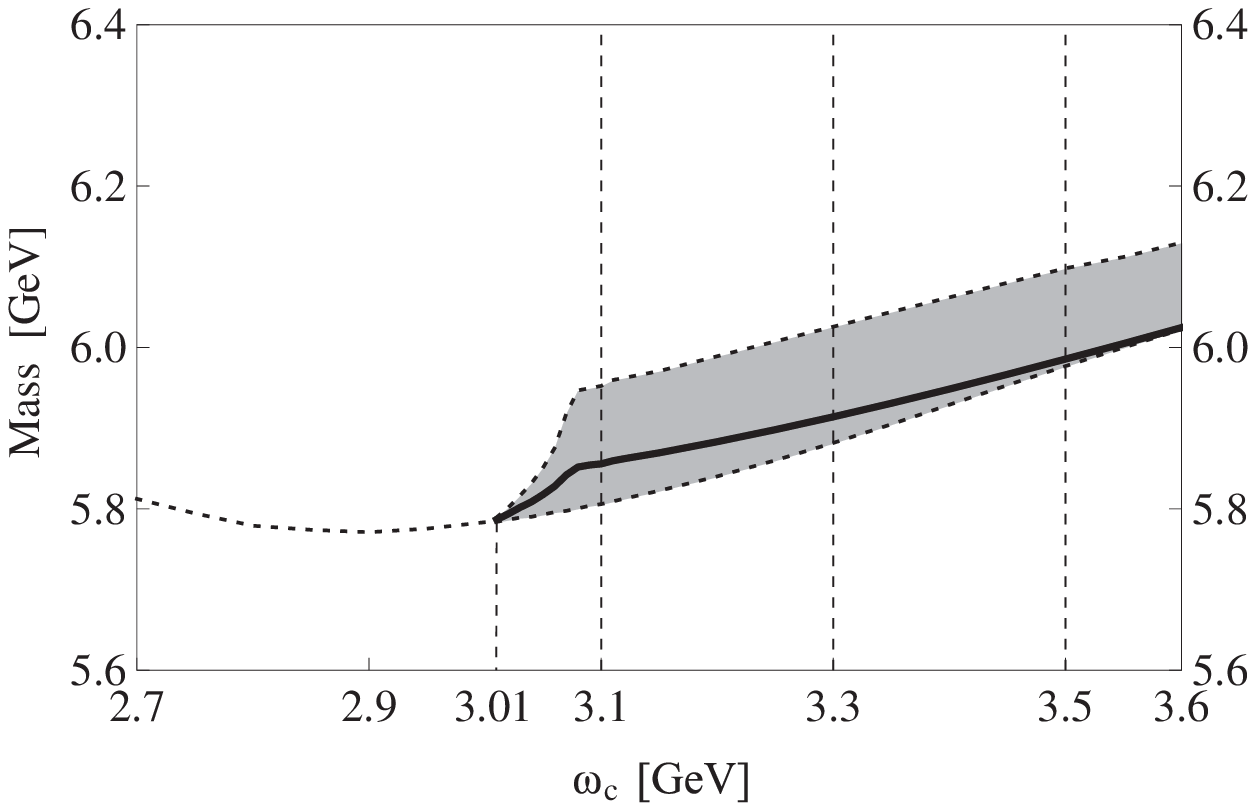}}
\end{tabular}
\caption{Variations of $m_{\Sigma_b(1/2^-)}$ with respect to the Borel mass
$T$ (top) and the threshold value $\omega_c$ (bottom), calculated using the bottom baryon doublet
$[\Sigma_b(\mathbf{6}_F), 1, 0, \rho]$. In the top panel we take $\omega_c = 3.3$ GeV and the Borel window is
$0.534$ GeV $< T < 0.624$ GeV.}
\label{fig:rho6F10}
\end{center}
\end{figure}

\item We use the bottom baryon doublet $[\Sigma_b(\mathbf{6}_F), 2, 1, \lambda]$ to perform QCD sum rule analyses. We show the variation of $m_{\Sigma_b(1/2^-)}$ with respect to the Borel mass $T$ and the threshold value $\omega_c$ in Fig.~\ref{fig:lambda6F21}. We find that there are non-vanishing Borel windows as long as $\omega_c \geq 2.97$ GeV, and the $\omega_c$ dependence is weak and acceptable in the region $3.1$ GeV$<\omega_c<3.5$ GeV. Accordingly, we choose $3.1$ GeV $<\omega_c <3.5$ GeV and $0.537$ GeV $< T < 0.620$ GeV as our working regions, and obtain the following numerical results
    \begin{eqnarray}
    \nonumber m_{\Sigma_b(3/2^-)} &=& 5.96 \pm 0.18 \mbox{ GeV} \, , \,
    \\ m_{\Sigma_b(5/2^-)} &=& 5.98 \pm 0.18 \mbox{ GeV} \, , \,
    \\ \nonumber \Delta m_{[\Sigma_b, 2, 1, \lambda]} &=& 15 \pm 7 \mbox{ MeV} \, ,
    \end{eqnarray}
    whose central values correspond to $T=0.579$ GeV and $\omega_c = 3.3$ GeV. We note again that the mass splitting within the same doublet is reproduced quite well with much less (theoretical) uncertainty.

\begin{figure}[hbt]
\begin{center}
\begin{tabular}{c}
\scalebox{0.6}{\includegraphics{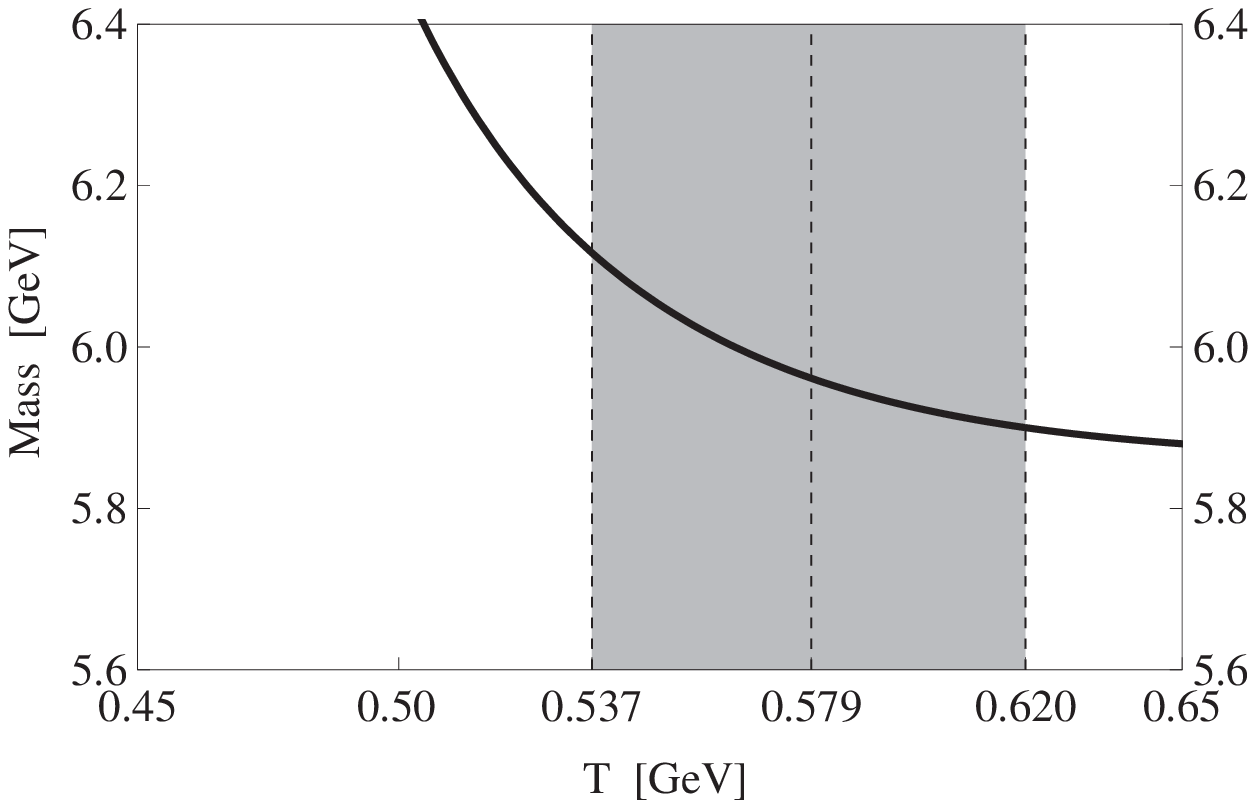}}
\\
\scalebox{0.6}{\includegraphics{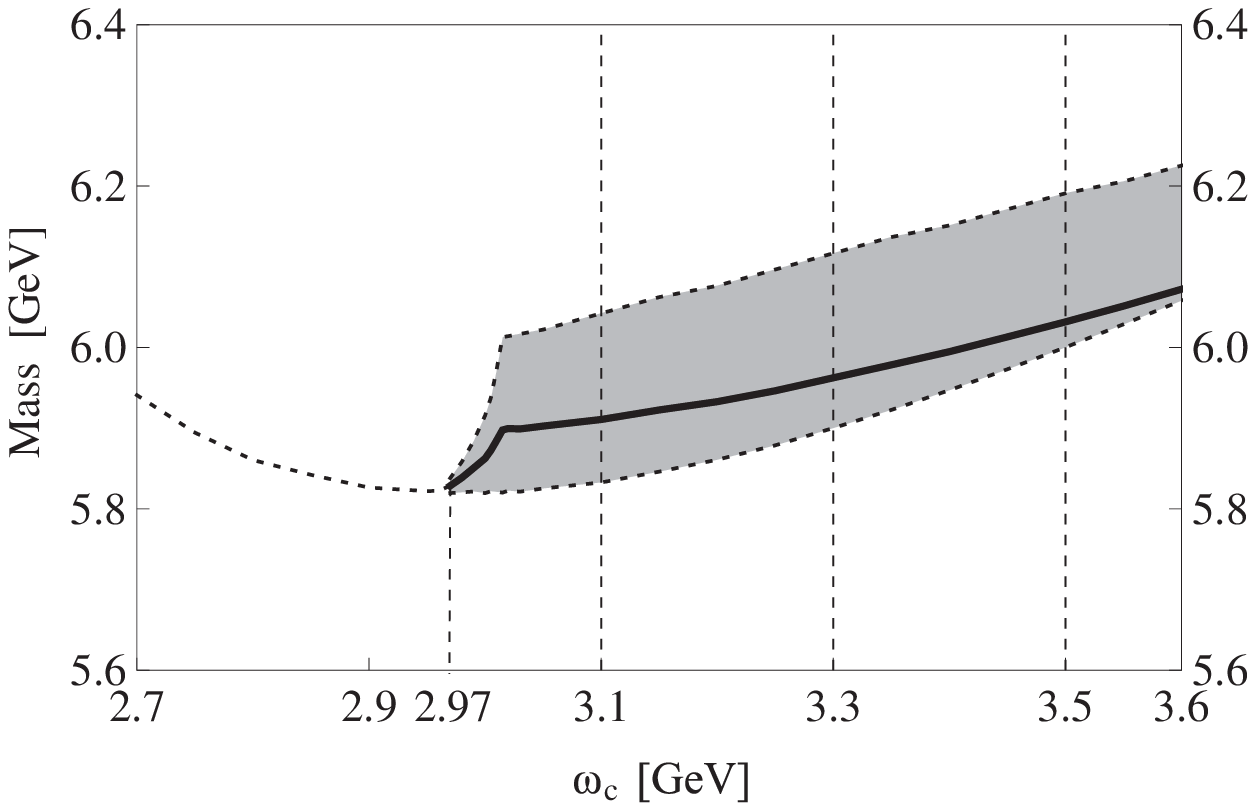}}
\end{tabular}
\caption{Variations of $m_{\Sigma_b(3/2^-)}$ with respect to the Borel mass
$T$ (top) and the threshold value $\omega_c$ (bottom), calculated using the bottom baryon doublet
$[\Sigma_b(\mathbf{6}_F), 2, 1, \lambda]$. In the top panel we take $\omega_c = 3.3$ GeV and the Borel window is
$0.537$ GeV $< T < 0.620$ GeV.}
\label{fig:lambda6F21}
\end{center}
\end{figure}

\item For flavor partners of $\Lambda_b$ and $\Sigma_b$, we use $\omega_c(\Xi_b)- \omega_c(\Lambda_b) = \omega_c(\Omega_b) - \omega_c(\Xi^\prime_b) = \omega_c(\Xi^\prime_b)- \omega_c(\Sigma_b)$ = 0.5 GeV, following the relations obtained/used for charmed baryons in Ref.~\cite{Chen:2015kpa}.

\item Other bottom baryon multiplets $[\mathbf{\bar 3}_F, 0, 1, \rho]$, $[\mathbf{\bar 3}_F, 2, 1, \rho]$, and $[\mathbf{6}_F, 1, 1, \lambda]$ do not give useful QCD sum rule results, which we shall not discuss to avoid confusions.

\end{enumerate}

We summarize these results obtained using the baryon multiplets $[\mathbf{\bar 3}_F, 1, 1, \rho]$, $[\mathbf{\bar 3}_F, 1, 0, \lambda]$, $[\mathbf{6}_F, 0, 1, \lambda]$, $[\mathbf{6}_F, 1, 0, \rho]$ and $[\mathbf{6}_F, 2, 1, \lambda]$ in Table~\ref{tab:results1}.

\begin{table*}[hbt]
\begin{center}
\caption{QCD sum rule results obtained using the baryon multiplets $[\mathbf{\bar 3}_F, 1, 1, \rho]$, $[\mathbf{\bar 3}_F, 1, 0, \lambda]$, $[\mathbf{6}_F, 0, 1, \lambda]$, $[\mathbf{6}_F, 1, 0, \rho]$ and $[\mathbf{6}_F, 2, 1, \lambda]$.}
\begin{tabular}{c | c | c | c | c | c c | c c | c}
\hline\hline
\multirow{2}{*}{Multiplets} & \multirow{2}{*}{B} & $\omega_c$ & Working region & $\overline{\Lambda}$ & $K$ & $\Sigma$ & Baryons & Mass & Difference
\\ & & (GeV) & (GeV) & (GeV) & (GeV$^2$) & (GeV$^2$) & ($j^P$) & (GeV) & (MeV)
\\ \hline\hline
\multirow{4}{*}{$[\mathbf{\bar 3}_F, 1, 1, \rho]$} & \multirow{2}{*}{$\Lambda_b$} & \multirow{2}{*}{3.1} & \multirow{2}{*}{$0.538< T < 0.588$} & \multirow{2}{*}{$1.17$} & \multirow{2}{*}{$-0.99$} & \multirow{2}{*}{$0.042$}
& $\Lambda_b(1/2^-)$ & $5.87 \pm 0.12$ & \multirow{2}{*}{$11 \pm 4$}
\\ \cline{8-9}
& & & & & & & $\Lambda_b(3/2^-)$ & $5.88 \pm 0.11$ &
\\ \cline{2-10}
& \multirow{2}{*}{$\Xi_b$} & \multirow{2}{*}{3.6} & \multirow{2}{*}{$0.536< T < 0.636$} & \multirow{2}{*}{$1.35$} & \multirow{2}{*}{$-0.98$} & \multirow{2}{*}{$0.035$}
& $\Xi_b(1/2^-)$ & $6.06 \pm 0.13$ & \multirow{2}{*}{$9 \pm 4$}
\\ \cline{8-9}
& & & & & & & $\Xi_b(3/2^-)$ & $6.07 \pm 0.13$ &
\\ \hline
\multirow{4}{*}{$[\mathbf{\bar 3}_F, 1, 0, \lambda]$} & \multirow{2}{*}{$\Lambda_b$} & \multirow{2}{*}{3.6} & \multirow{2}{*}{$0.597< T < 0.684$}
& \multirow{2}{*}{$1.21$} & \multirow{2}{*}{$-2.60$} & \multirow{2}{*}{$0.019$} & $\Lambda_b(1/2^-)$ & $6.01 \pm 0.11$ & \multirow{2}{*}{$5 \pm 2$}
\\ \cline{8-9}
& & & & & & & $\Lambda_b(3/2^-)$ & $6.01 \pm 0.11$ &
\\ \cline{2-10}
& \multirow{2}{*}{$\Xi_b$} & \multirow{2}{*}{4.1} &
\multirow{2}{*}{$0.565< T < 0.759$}
& \multirow{2}{*}{$1.44$} & \multirow{2}{*}{$-3.21$} & \multirow{2}{*}{$0.014$} & $\Xi_b(1/2^-)$ & $6.27 \pm 0.13$ & \multirow{2}{*}{$4 \pm 2$}
\\ \cline{8-9}
& & & & & & & $\Xi_b(3/2^-)$ & $6.28 \pm 0.13$ &
\\ \hline
\multirow{3}{*}{$[\mathbf{6}_F, 0, 1, \lambda]$} & $\Sigma_b$ & 3.3 &
{$0.527< T < 0.628$}
& $1.22$ & $-2.53$ & $0$ & $\Sigma_b(1/2^-)$ & $6.02 \pm 0.12$ & $-$
\\ \cline{2-10}
& $\Xi^\prime_b$ & 3.8 & {$0.531< T < 0.676$} & $1.43$ & $-2.77$ & $0$ & $\Xi^\prime_b(1/2^-)$ & $6.24 \pm 0.11$ & $-$
\\ \cline{2-10}
& $\Omega_b$ & 4.3 & {$0.547< T < 0.740$} & $1.67$ & $-3.22$ & $0$ & $\Omega_b(1/2^-)$ & $6.50 \pm 0.11$ & $-$
\\ \hline
\multirow{6}{*}{$[\mathbf{6}_F, 1, 0, \rho]$} & \multirow{2}{*}{$\Sigma_b$} & \multirow{2}{*}{3.3} & \multirow{2}{*}{$0.534< T < 0.624$} & \multirow{2}{*}{$1.19$} & \multirow{2}{*}{$-1.16$} & \multirow{2}{*}{$0.014$} & $\Sigma_b(1/2^-)$ & $5.91 \pm 0.14$ & \multirow{2}{*}{$4 \pm 2$}
\\ \cline{8-9}
& & & & & & & $\Lambda_b(3/2^-)$ & $5.92 \pm 0.14$ &
\\ \cline{2-10}
& \multirow{2}{*}{$\Xi^\prime_b$} & \multirow{2}{*}{3.8} & \multirow{2}{*}{$0.521< T < 0.681$} & \multirow{2}{*}{$1.38$} & \multirow{2}{*}{$-1.27$} & \multirow{2}{*}{$0.011$}
& $\Xi^\prime_b(1/2^-)$ & $6.11 \pm 0.13$ & \multirow{2}{*}{$3 \pm 1$}
\\ \cline{8-9}
& & & & & & & $\Xi^\prime_b(3/2^-)$ & $6.11 \pm 0.13$ &
\\ \cline{2-10}
& \multirow{2}{*}{$\Omega_b$} & \multirow{2}{*}{4.3} & \multirow{2}{*}{$0.510< T < 0.751$} & \multirow{2}{*}{$1.60$} & \multirow{2}{*}{$-1.56$} & \multirow{2}{*}{$0.009$}
& $\Omega_b(1/2^-)$ & $6.34 \pm 0.13$ & \multirow{2}{*}{$2 \pm 1$}
\\ \cline{8-9}
& & & & & & & $\Omega_b(3/2^-)$ & $6.34 \pm 0.13$ &
\\ \hline
\multirow{6}{*}{$[\mathbf{6}_F, 2, 1, \lambda]$} & \multirow{2}{*}{$\Sigma_b$} & \multirow{2}{*}{3.3} & \multirow{2}{*}{$0.537< T < 0.620$} & \multirow{2}{*}{$1.17$} & \multirow{2}{*}{$-2.58$} & \multirow{2}{*}{$0.035$}
& $\Sigma_b(3/2^-)$ & $5.96 \pm 0.18$ & \multirow{2}{*}{$15 \pm 7$}
\\ \cline{8-9}
& & & & & & & $\Sigma_b(5/2^-)$ & $5.98 \pm 0.18$ &
\\ \cline{2-10}
& \multirow{2}{*}{$\Xi^\prime_b$} & \multirow{2}{*}{3.8} & \multirow{2}{*}{$0.525< T < 0.684$} & \multirow{2}{*}{$1.36$} & \multirow{2}{*}{$-2.90$} & \multirow{2}{*}{$0.028$} & $\Xi^\prime_b(3/2^-)$ & $6.17 \pm 0.17$ & \multirow{2}{*}{$12 \pm 6$}
\\ \cline{8-9}
& & & & & & & $\Xi^\prime_b(5/2^-)$ & $6.18 \pm 0.16$ &
\\ \cline{2-10}
& \multirow{2}{*}{$\Omega_b$} & \multirow{2}{*}{4.3} & \multirow{2}{*}{$0.518< T < 0.759$} & \multirow{2}{*}{$1.59$} & \multirow{2}{*}{$-3.45$} & \multirow{2}{*}{$0.022$} & $\Omega_b(3/2^-)$ & $6.43 \pm 0.13$ & \multirow{2}{*}{$9 \pm 4$}
\\ \cline{8-9}
& & & & & & & $\Omega_b(5/2^-)$ & $6.43 \pm 0.13$ &
\\ \hline \hline
\end{tabular}
\label{tab:results1}
\end{center}
\end{table*}

\section{Discussion and Conclusion}
\label{sec:summary}

In this paper we followed Ref.~\cite{Chen:2015kpa} and studied the $P$-wave bottom baryons using the method of QCD sum rule in the framework of HQET. We note that we just need to redo numerical analyses using the $bottom$ quark mass. The results are summarized in Table~\ref{tab:results1} for baryon multiplets $[\mathbf{\bar 3}_F, 1, 1, \rho]$, $[\mathbf{\bar 3}_F, 1, 0, \lambda]$, $[\mathbf{6}_F, 0, 1, \lambda]$, $[\mathbf{6}_F, 1, 0, \rho]$ and $[\mathbf{6}_F, 2, 1, \lambda]$.

Our results suggest that the two observed states $\Lambda_b(5912)^0$ ($J^P=1/2^-$) and $\Lambda_b(5920)^0$ ($J^P=3/2^-$)~\cite{Aaij:2012da,Aaltonen:2013tta} can be well described by the baryon doublet $[\mathbf{\bar 3}_F, 1, 1, \rho]$. They are the bottom partners of $\Lambda_c(2595)$ and $\Lambda_c(2625)$~\cite{pdg}, and also belong to the $SU(3)$ $\mathbf{\bar 3}_F$ multiplets of $J^P=1/2^-$ and $3/2^-$. Our results suggest that their $SU(3)$ flavor partners, $\Xi_b(1/2^-)$ and $\Xi_b(3/2^-)$, have masses $6.06 \pm 0.13$ GeV and $6.07 \pm 0.13$ GeV, respectively, with a mass splitting $9 \pm 4$ MeV.

The QCD sum rule results obtained by using baryon doublet $[\mathbf{\bar 3}_F, 1, 0, \lambda]$ are similar and also consistent with the experimental data~\cite{Aaij:2012da,Aaltonen:2013tta}. Here we obtain that the $\rho$ mode excitation, $[\mathbf{\bar 3}_F, 1, 1, \rho]$, is lower than the $\lambda$ one, $[\mathbf{\bar 3}_F, 1, 0, \lambda]$. This is consistent with our previous analysis of the charm sector~\cite{Chen:2015kpa}, in contrast with the quark model expectation~\cite{Copley:1979wj,Yoshida:2015tia}. Thus, further studies are needed to clarify their nature. Particularly, it would be useful to study decays and productions of these heavy baryon excitations.

We also studied the $SU(3)$ $\mathbf{6}_F$ multiplets by using the baryon multiplets $[\mathbf{6}_F, 0, 1, \lambda]$, $[\mathbf{6}_F, 1, 0, \rho]$ and $[\mathbf{6}_F, 2, 1, \lambda]$. Our results suggest that the $P$-wave bottom baryons $\Sigma_b$, $\Xi^\prime_b$ and $\Omega_b$ have (averaged) masses about 6.0 GeV, 6.2 GeV and 6.4 GeV, respectively.

While checking the status of observed bottom baryons listed in PDG, we notice that progress have been made in the experimental observations. Especially, in the past years, experiments like LHCb and CDF announced the observations of higher radial excitations of bottom baryon~\cite{Aaij:2012da,Aaltonen:2013tta}. We believe that higher radial and orbital excitations will be released in future experiments. With the running of LHCb at 13 TeV, experimental exploration to higher radial and orbital excitations of bottom baryons will be a penitential and intriguing research topic. The present theoretical results of the mass of $P$-wave bottom baryons can provide valuable information. In near future, the joint efforts from experimentalist and theorist will be helpful to identify more and more bottom baryons.

\section*{ACKNOWLEDGMENTS}

We thank Cheng-Ping Shen and Ke-Wei Wei for useful discussions and careful reading of the manuscript.
H.X.C. thanks Profs. Luciano Maiani, Rinaldo Baldini and Pierluigi Campana
for their hospitality during his stay at INFN-LNF.
This project is supported by the National Natural Science Foundation
of China under Grants No. 11205011, No. 11475015, No. 11375024, No. 11222547, No.
11175073, and No. 11261130311, the Ministry of
Education of China (SRFDP under Grant No. 20120211110002 and the
Fundamental Research Funds for the Central Universities).

\end{document}